\documentclass[aip,jcp,reprint,groupedaddress,floatfix]{revtex4-1}
\usepackage{graphicx}
\usepackage[english]{babel}
\usepackage{amsmath}
\usepackage{amssymb}
\usepackage{dcolumn}
\usepackage{color}
\usepackage{comment}
\usepackage{braket}
\includecomment{reprint}
\excludecomment{preprint}

\renewcommand{\Re}{\operatorname{Re}}
\renewcommand{\Im}{\operatorname{Im}}
\newcommand{\rrangle}{\rangle\!\rangle}
\newcommand{\llangle}{\langle\!\langle}
\begin{document}

\title{Symplectic integration and physical interpretation of time-dependent coupled-cluster theory}
\author{Thomas Bondo Pedersen}
\email{t.b.pedersen@kjemi.uio.no}
\affiliation{Hylleraas Centre for Quantum Molecular Sciences,
Department of Chemistry, University of Oslo,
P.O. Box 1033 Blindern, N-0315 Oslo, Norway}
\author{Simen Kvaal}
\email{simen.kvaal@kjemi.uio.no}
\affiliation{Hylleraas Centre for Quantum Molecular Sciences,
Department of Chemistry, University of Oslo,
P.O. Box 1033 Blindern, N-0315 Oslo, Norway}

\date{\today}

\begin{abstract}
  The formulation of the time-dependent Schr\"odinger equation in
  terms of coupled-cluster theory is outlined, with emphasis on the
  bivariational framework and its classical Hamiltonian structure. An
  indefinite inner product is introduced, inducing physical
  interpretation of coupled-cluster states in the form of transition
  probabilities, autocorrelation functions, and explicitly real values
  for observables, solving interpretation issues which are present in
  time-dependent coupled-cluster theory and in ground-state calculations
  of molecular systems under influence of external magnetic
  fields. The problem of the numerical integration of the equations of
  motion is considered, and a critial evaluation of the standard
  fourth-order Runge--Kutta scheme  and the symplectic Gauss integrator
  of variable order is given, including several illustrative numerical
  experiments. While the Gauss integrator is stable even for laser pulses
  well above the perturbation limit, our experiments indicate that a system-dependent
  upper limit exists for the external field strengths. Above this limit,
  time-dependent coupled-cluster calculations become very challenging numerically,
  even in the full configuration interaction limit.
  The source of these numerical instabilities is shown to be rapid increases of the
  amplitudes as ultrashort high-intensity laser pulses pump the system
  out of the ground state into states that are virtually orthogonal to the
  static Hartree-Fock reference determinant.
\end{abstract} 

\maketitle

\section{Introduction}

Originally developed as a description of short-range interactions in
the ground state of closed-shell atomic nuclei,~\cite{Coester1960}
the coupled-cluster (CC) method has evolved into the most reliable wave
function-based computational tool in quantum
chemistry. It is routinely
applied to molecular electronic ground- and excited-state energies,
structures, and properties---see Refs.~\onlinecite{Bartlett2007}
and~\onlinecite{Helgaker2012} for reviews. These developments have
been based mainly on the CC wave function as an \emph{Ansatz} for
solving the time-independent Schr{\"o}dinger equation. Although
time-dependent CC theory, which also has its roots in nuclear
physics,~\cite{Hoodbhoy1978,Hoodbhoy1979} forms the starting point for
a perturbative description of frequency-dependent response
properties,~\cite{Helgaker2012,Koch1990} it has only rarely been used
for the study of real-time many-electron dynamics.

One can imagine several reasons for the lack of interest in explicitly
time-dependent CC theory, one being the anticipated steep increase in
computational cost compared with the calculation of ground- and
excited-state energies. More serious-sounding is perhaps the tendency
for observables to acquire nonzero imaginary parts, which implies that
the \emph{interpretation} of time-dependent CC calculations is
non-trivial. Indeed, there seems to be some disagreement on how to
interpret the coupled-cluster state. This problem stems from the
non-variational nature of CC theory, and should show up whenever
ground- and excited-state calculations are carried out on \emph{complex}
Hamiltonians, such as molecular systems in external magnetic
fields.~\cite{Stopkowicz2015,Hampe2017}

Even if CC theory is nonvariational and significantly more
expensive than the indisputable work-horse of electronic-structure
theory, Kohn-Sham density-functional
theory,~\cite{Hohenberg1964,Kohn1965} for the calculation of the same
quantities, development of increasingly sophisticated CC
methods has continued to this day. Recent algorithmic advances have
made highly accurate CC calculations of ground-state energies a nearly
routine endeavor, see, e.g.,
Refs.~\onlinecite{Eriksen2015,Schwilk2017,Pavosevic2017}. 

Another likely reason for the lack of interest
in explicitly time-dependent CC theory is lack of scientific
imperative.  The majority of interesting chemical problems involve
only the electronic ground state and, in some cases, perhaps a few
excited states. In addition, even sophisticated higher-order
spectroscopies can be adequately described using response theory,
making an explicitly time-dependent treatment unnecessary.

As illustrated by the 2018 Nobel Prize in Physics,~\cite{NP2018} the
situation has changed dramatically in recent years due to
breakthroughs in the generation of high-intensity, ultrashort laser
pulses.~\cite{Krausz2009} Such pulses create an extreme environment for
the particle dynamics, violating the basic assumptions of the
perturbation theory that underpins response theory, and thus forcing
us to focus on solving the time-dependent Schr{\"o}dinger equation
directly. A spatial high-resolution description of a many-electron
wavefunction is hugely expensive, scaling exponentially with the
number of electrons.  Hence, the standard way to formulate
time-dependent electronic wavefunctions today is the
multi-configurational time-dependent Hartree--Fock method (MCTDHF),
see Ref.~\onlinecite{Meyer2009} and references therein. More
generally, the time-dependent complete active space self-consistent
field method (TD-CASSCF),~\cite{Sato2013} along with the
restricted active space (TD-RASSCF)~\cite{Miyagi2013,Hochstuhl2014}
and generalized active space (TD-GASSCF)~\cite{Bauch2014}
versions, have
been developed to reduce the cost of ionization simulations by means
of their active space formulation. However, the exponential cost of the
wavefunction is only delayed with such methods, making the application
to larger systems too expensive.

Considering that the correlation description of CC theory is only \emph{polynomially}
scaling with the number of electrons, it should be an
interesting candidate for high-accuracy simulations in this
area. Despite this fact, very few studies of time-dependent CC theory
have been performed. 

Sch{\"o}nhammer and Gunnarsson~\cite{Schonhammer1978} used
time-dependent CC singles-and-doubles (CCSD) to compute the spectral
function of an approximate many-body Hamiltonian with respect to the
Hartree-Fock (HF) determinant and applied it to the study of
photoemission from adsorbate core levels.  More recently, Huber and
Klamroth~\cite{Huber2011} studied laser-driven many-electron dynamics
in small closed-shell molecules at the time-dependent CCSD level,
using the explicit fourth-order Runge-Kutta (RK4) integrator to
propagate the CCSD amplitudes and computing the induced dipole moment
as a function of time through a configuration-interaction
singles-and-doubles (CISD) wave function constructed from the CCSD
amplitudes in each time step.  They made the rather worrying
observation that the use of Gaussian basis sets larger than Pople's
6-31G* basis~\cite{Hehre1972} and/or increasing the field strength
beyond about $10^{-3}\,\text{au}$ led to numerical instabilities in
the integration.

In order to describe ionization accurately, Kvaal~\cite{Kvaal2012}
suggested a time-dependent orbital-adaptive CC method (OACC), a
theoretical framework that gives a hierarchy of methods that
interpolate between time-dependent Hartree--Fock on one end, and
MCTDHF on the other end. The method is 
similar to the time-dependent nonorthogonal orbital-optimized CC
(NOCC) method of Pedersen et al.~\cite{Pedersen2001} in the sense that
the orbitals and amplitudes are determined in a concerted fashion,
thus avoiding spurious uncorrelated resonances. Kvaal used a
variational splitting scheme,~\cite{Lubich2004} effectively pulsing the
electronic interaction, to
improve stability of the RK4 method through a near-exact description of high-frequency
oscillatory amplitude components. Recently, Sato et
al.~\cite{Sato2018} employed a similar orbital-adaptive CC theory,
orbital-optimized CC (OCC),~\cite{Sherrill1998,Pedersen1999,Krylov2000}
which differs from OACC and NOCC by
enforcing orthonormality of the orbitals, to study higher harmonic
generation and one- and two-electron ionization of the Ar atom in an
intense laser pulse. A formal problem of the OCC theory is that it
does not converge to the exact full configuration-interaction (FCI)
limit for systems with more than two electrons whereas NOCC (and OACC)
theory does.~\cite{Kohn2005,Myhre2018} Sato et al. used an exponential
RK4 integrator, presumably to avoid instabilities arising from highly
oscillatory cluster amplitudes.

Kvaal's work,~\cite{Kvaal2012} which is based on Arponen's bivariational formulation,~\cite{Arponen1983}
inspired renewed efforts by Pigg et al.~\cite{Pigg2012} to study nucleon dynamics using time-dependent CC theory.
Their focus was on ground- and excited-state energies, the latter obtained by Fourier transformation of randomly
selected individual singles and doubles amplitudes, and they explicitly demonstrated that observables that
commute with the Hamiltonian are conserved under exact propagation. Upon discretization, they found that total energy variation
ranged from insignificant to substantial depending on the time step used in the RK4 integrator.

Nascimento and DePrince
proposed a time-dependent extension of equation-of-motion CC (EOM-CC)~\cite{Stanton1993,Krylov2008}
theory with the somewhat reduced scope, compared with the papers cited above, of computing linear absorption
spectra.~\cite{Nascimento2016} This was achieved by combining Fermi's Golden Rule, a result from first-order perturbation theory,
with the EOM-CC parameterization. They subsequently used this formulation to simulate near-edge x-ray fine structure.~\cite{Nascimento2017}

In this work we consider the conventional CC theory constructed on top
of a static HF determinant.  Particular to CC theory is that it
is best cast in a bivariational framework as pioneered by Arponen and
coworkers,~\cite{Arponen1983} meaning that it is variational in a
generalized sense, with the expectation value functional generated by
\emph{variationally independent} bra and ket wavefunctions. Notably,
as both bra and ket functions are needed to represent the quantum
mechanical state, it is not physically meaningful to talk about one
without the other, and by symmetry the bra and the ket should be
treated on equal footing. We outline the bivariational theory,
and address the interpretation issue by introducing an indefinite inner
product that induces expectation values, autocorrelation functions,
transition amplitudes, and therefore most of the formalism needed to
study the physics of the time evolution. In particular, all physical
quantities are manifestly real. 

We also consider the problem of choosing a suitable numerical
integrator for the equations of motion.
As seen from the brief survey above, the most commonly used
integrator for time-dependent CC theory is the explicit RK4 integrator,
presumably because of it's ease of implemenation and relatively low
computational cost. It is not unlikely, however, that other 
explicit integrators are more efficient, such as the Bulirsch-Stoer
scheme~\cite{Stoer1993} which was successfully applied to 
time-dependent algebraic-diagrammatic construction (ADC)
theory by Neville and Schuurman.~\cite{Neville2018}
In exact quantum theory the
time-dependent Schr\"odinger equation can be formulated as an abstract
Hamiltonian mechanical system. The bra and the ket form a point in
phase space, which is infinite-dimensional.~\cite{Chernoff1974}
This classical Hamiltonian structure is preserved
with an approximate finite-dimensional linear parameterization
of the wave function, the real and imaginary parts of the parameters serving as
generalized coordinates and conjugate momenta, respectively, see for example
Ref.~\onlinecite{Gray1996}.
The CC nonlinear parameterization in terms of cluster amplitudes is canonical,
preserving the structure of Hamilton's equations
in complex form.
The coordinates are
the usual exponentially occurring amplitudes, while the momenta are
the Lagrange multipliers. The Hamilton function in this formulation
is, somewhat confusingly, the conventional CC Lagrangian. Looking for
a numerically stable integrator for the time-dependent CC equations,
these observations allow us to draw upon the vast numerical experience with
classical Hamiltonian systems.

This paper is organized as follows. In Sec.~\ref{sec:theory} we discuss
the Hamiltonian structure of time-dependent CC theory in more detail,
propose an indefinite inner product and corresponding autocorrelation
function for analysis of the many-electron  
dynamics, and describe a suitable symplectic integrator.
Numerical experiments are presented in Sec.~\ref{sec:numexp},
including high-intensity laser pulses, and concluding remarks are
given in Sec.~\ref{sec:conclusion}

\section{Theory}
\label{sec:theory}

\subsection{Time-dependent coupled-cluster equations}

Given a time-dependent electronic Hamiltonian $H(t)$, the starting
point is an action-like functional introduced by
Arponen~\cite{Arponen1983} (but see also Chernoff and Marsden~\cite{Chernoff1974}),
\begin{equation}
   {\cal S}[\bra{\tilde{\Psi}},\ket{\Psi}] = \int_0^T \langle \tilde{\Psi}(t) \vert
   \mathrm{i} \frac{\mathrm{d}}{\mathrm{d}t} - H(t) 
                         \vert \Psi(t) \rangle\,
                         \text{d}t. \label{eq:action}
\end{equation}
The stationary points with respect to variations in the bra and the
ket vectors are, respectively, the time-dependent Schr\"odinger
equation and its dual. Letting the bra and the ket be
momenta and coordinates, respectively, in an infinite-dimensional phase
space, we see that the stationary condition is
nothing but Hamilton's modified principle\cite{Goldstein_CM} (with the
appearance of an additional imaginary unit) for the Hamilton
function $\mathcal{H} = \braket{\tilde{\Psi}(t)|H(t)|\Psi(t)}$,
\begin{subequations} \label{eq:schroedinger}
\begin{align}
  \mathrm{i} \ket{\dot{\Psi}(t)} &= \frac{\partial\mathcal{H}}{\partial \bra{\tilde{\Psi}(t)}},
\\
  \mathrm{i} \bra{\dot{\tilde\Psi}(t)} &= -\frac{\partial\mathcal{H}}{\partial \ket{\Psi(t)}},
\end{align}
\end{subequations}
where the dot denotes the time derivative.
We parameterize the bra and the
ket vectors relative to a static reference Slater determinant $\vert \Phi_0 \rangle$---in practice,
the HF ground-state determinant at time $t=0$---as
\begin{subequations}
\begin{align}
\label{eq:tPsi}
   &\langle \tilde{\Psi}(t) \vert = \langle \Phi_0 \vert \bar{\Lambda}(t) \exp(-\bar{T}(t)), \\
\label{eq:Psi}
   &\vert \Psi(t) \rangle = \exp(\bar{T}(t)) \vert \Phi_0 \rangle.
\end{align}
\end{subequations}
The cluster operators are defined in terms of particle-conserving excitation operators $X_\mu$ with respect to $\vert \Phi_0 \rangle$
and associated time-dependent amplitudes $\tau_\mu$ and $\lambda_\mu$
as
\begin{subequations}
\begin{align}
\label{eq:barT}
   &\bar{T}(t) = \sum_{\mu \geq 0} \tau_\mu(t) X_\mu, \\
\label{eq:barLambda}
   &\bar{\Lambda}(t) = \sum_{\mu \geq 0} Y_\mu^\dagger \lambda_\mu(t),
\end{align}
\end{subequations}
where both summations run over the same set and include at most $N$-electron excitations for an $N$-electron system.
The deexcitation operators $Y_\mu^\dagger$ are defined through an orthogonal transformation of the operators
$X_\mu^\dagger$ such that the biorthonormality condition
\begin{equation}
   \langle \Phi_0 \vert Y_\mu^\dagger X_\nu \vert \Phi_0 \rangle = \delta_{\mu\nu},
\end{equation}
is fulfilled.  Note that we use the convention $X_0 = Y_0 = I$, where
$I$ is the identity operator, such that phase and normalization of the
bra and ket vectors are determined by the amplitudes
$\tau_0$ and $\lambda_0$. The parameterization in terms of cluster
amplitudes is \emph{exact}, even in the full infinite-dimensional
case,~\cite{Rohwedder2013} for all $\ket{\Psi}$ such that
$\braket{\Phi_0|\Psi}\neq 0$ and all $\bra{\tilde{\Psi}}$ such that
$\braket{\tilde{\Psi}|\Psi} \neq 0$.

Writing the action functional in terms of the amplitudes and
omitting the time variable for clarity, we obtain
\begin{equation}
  \mathcal{S}[\lambda,\tau] = \int_0^T  \text{i} \lambda\cdot \dot{\tau} -
  \mathcal{H}(\tau,\lambda) \, \text{d}t,
\end{equation}
which exhibits the transformation from amplitudes to wavefunctions as a
canonical transformation in the sense of classical mechanics. Here
$\mathcal{H}(\lambda,\tau) =
\braket{\tilde{\Psi}(\tau,\lambda)|H|\Psi(\tau)}$, i.e., the
conventional CC Lagrangian (not to be confused with a
hypothetical Lagrangian in the sense of classical mechanics, which,
in fact, does not exist). In the exact case, $\mathcal{H}$ is the energy expectation
value if $\braket{\tilde{\Psi}|\Psi}=\lambda_0 = 1$. 

Requiring that ${\cal S}$ be stationary with respect to variations in the amplitudes
we obtain the ordinary differential
equations~\cite{Arponen1983,Pedersen1998}
\begin{equation}
\label{eq:Hamilton}
   \text{i}\dot{\tau}_\mu = \frac{\partial {\cal H}}{\partial \lambda_\mu},
   \qquad
   \text{i}\dot{\lambda}_\mu = -\frac{\partial {\cal H}}{\partial \tau_\mu},
   \qquad \mu \geq 0.
\end{equation}
Like Eq.~\eqref{eq:schroedinger}, these are complex but otherwise classical Hamiltonian
equations. They can be brough to standard real form by
  setting $\tau = (q_1 + ip_2)/\sqrt{2}$ and $\lambda = (q_2 - i
  p_1)/\sqrt{2}$, with the real part of $\mathcal{H}$ as Hamiltonian function.
Noting that ${\cal H}$ does not depend on the amplitude $\tau_0$, we may
write the Hamiltonian equations as
\begin{subequations}
\begin{align}
\label{eq:rhs1}
    \text{i}\dot{\tau}_0 &= \langle \Phi_0 \vert H \vert \psi \rangle,
  &  \text{i}\dot{\lambda}_0 & = 0, \\
\label{eq:rhs2}
   \text{i}\dot{\tau}_\mu &= \langle \Phi_\mu \vert \exp(-T)H \vert \psi \rangle, &
     \text{i}\dot{\lambda}_\mu &= -\langle \tilde{\psi} \vert [H, X_\mu] \vert \psi \rangle,
\end{align}
\end{subequations}
where $\mu>0$ and
$\langle \Phi_\mu \vert = \langle \Phi_0 \vert Y_\mu^\dagger$. The
amplitude $\lambda_0$ is time-independent and may be fixed once and
for all to $\lambda_0=1$. The evolution of $\tau_0$ is decoupled from
the other amplitudes, and it is thus convenient to separate out the
normalization amplitudes and define
\begin{subequations}
\begin{align}
   &\langle \tilde{\Psi}(t) \vert = \exp(-\tau_0(t)) \langle
     \tilde{\psi}(t) \vert, \\
   &\vert \Psi(t) \rangle = \vert \psi(t) \rangle \exp(\tau_0(t)),
\end{align}
\end{subequations}
where
\begin{subequations}
\begin{align}
\label{eq:tpsi}
   &\langle \tilde{\psi}(t) \vert = \langle \Phi_0 \vert (1 + \Lambda(t)) \exp(-T(t)), \\
\label{eq:psi}
   &\vert \psi(t) \rangle = \exp(T(t)) \vert \Phi_0 \rangle, \\
\label{eq:T}
   &T(t) = \sum_{\mu > 0} \tau_\mu X_\mu, \\
\label{eq:L}
   &\Lambda(t) = \sum_{\mu > 0} \lambda_\mu Y_\mu.
\end{align}
\end{subequations}
This parameterization corresponds to the time-dependent CC states of
Koch and J{\o}rgensen,~\cite{Koch1990} who derived frequency-dependent
linear and quadratic response functions through a time-dependent
extension of the general Lagrangian formulation of static molecular
properties by Helgaker and
J{\o}rgensen.~\cite{Jorgensen1988,Helgaker1989,Helgaker1992} The
relation to extended CC theory has been discussed in detail by Arponen
et al.~\cite{Arponen1983,Arponen1987,Arponen1987a} 

We note in passing that, with
appropriate initial conditions, the phase factor $\exp (\tau_0(t))$ is
related to the spectral weight function of the Hamiltonian with
respect to the HF determinant, as pointed out by Sch{\"o}nhammer and
Gunnarsson,~\cite{Schonhammer1978} who used it to study x-ray
photoemission from adsorbate core levels.

\subsection{Expectation values and autocorrelation functions}

As CC theory is not variational in the usual sense, both $\vert
\Psi\rangle$ and $\bra{\tilde{\Psi}}$ at an instant in time are
required in order to fully represent a quantum state. In order to have
a balanced treatment of the two, we define a state vector
\begin{equation}
  | S \rrangle = \frac{1}{\sqrt{2}}
   \begin{pmatrix}
      \vert \Psi\rangle \\
      \vert \tilde{\Psi}\rangle
   \end{pmatrix},
\label{eq:S}
\end{equation}
for which we define the \emph{indefinite} inner product
\begin{equation}
  \begin{split}
   \llangle S_1 \vert S_2\rrangle
   &\equiv \frac{1}{2} \left( \langle \tilde{\Psi}_1 \vert \Psi_2 \rangle
                  + \langle \Psi_1 \vert \tilde{\Psi}_2 \rangle \right)
   \\
\label{eq:iip}
   &= \frac{1}{2} \left( \langle \tilde{\Psi}_1 \vert \Psi_2 \rangle
                  + \langle \tilde{\Psi}_2 \vert \Psi_1 \rangle^* \right).
\end{split}
\end{equation}
We use this scalar product to define transition amplitudes and
expectation values, and note that $|S(t)\rrangle$ is normalized with respect
to this inner product for all $t$ provided $\lambda_0 = 1$.

The expectation value of an operator $P$ may then be computed with respect to the indefinite inner product
as
\begin{equation}
  \llangle S \vert \hat{P} \vert S\rrangle
   = \frac{1}{2} \langle \tilde{\Psi} \vert P \vert \Psi\rangle
   + \frac{1}{2} \langle \tilde{\Psi} \vert P^\dagger \vert \Psi\rangle^*,
\end{equation}
where
\begin{equation}
   \hat{P} =
   \begin{pmatrix}
      P & 0 \\
      0 & P
   \end{pmatrix}.
\end{equation}
Thus defined, the expectation value satisfies the Hellmann-Feynman theorem~\cite{Arponen1983,Arponen1987,Pedersen1998}
and whenever $P$ is Hermitian, 
\begin{equation}
  \llangle S \vert \hat{P} \vert S \rrangle = \Re \mathcal{P}, \quad
  \mathcal{P} \equiv \braket{\tilde{\Psi}|P|\Psi},
\end{equation}
producing real values for Hermitian operators. The latter is important
to, e.g., ensure proper permutation symmetries of response functions such as the electric dipole
polarizability.~\cite{Pedersen1997} 

We note that the indefinite inner product induces an action functional
which is equivalent to $\mathcal{S}$ whenever $H$ is
Hermitian. Indeed, $\int_0^T \llangle S(t)|(\mathrm{i}\partial_{t} -
\hat{H})|S(t)\rrangle\, \mathrm{d}t = \Re \mathcal{S}[\bra{\tilde{\Psi}},\ket{\Psi}]$. Since $\mathcal{S}$
is complex differentiable, the two actions have the same critical points.

It follows from Eq.~\eqref{eq:Hamilton} that the amplitudes $\tau_\mu,\lambda_\mu\ (\mu \geq 0)$ are canonical variables 
defining a phase space analogous to generalized position and momentum variables in classical Hamiltonian
mechanics.~\cite{Arponen1987,Pedersen1998}
Introducing the generalized Poisson bracket
\begin{equation}
   \{ {\cal P}, {\cal Q} \} \equiv
   \sum_{\mu \geq 0}
   \left(
       \frac{\partial {\cal P}}{\partial \tau_\mu} \frac{\partial {\cal Q}}{\partial \lambda_\mu}
       -
       \frac{\partial {\cal Q}}{\partial \tau_\mu} \frac{\partial {\cal P}}{\partial \lambda_\mu}
   \right),
\end{equation}
we find that the time evolution of the expectation-value function obeys
\begin{equation}
   \text{i}\dot{{\cal P}} = \{ {\cal P}, {\cal H} \} + \text{i}\frac{\partial {\cal P}}{\partial t},
\end{equation}
where the last term is relevant only if the operator $P$ is explicitly time-dependent (the last term is the expectation
value of the time-derivative of $P$). This allows us to identify conservation laws even with truncated cluster
operators. In particular, of course, energy is conserved when the Hamiltonian operator $H$ is time-independent,
$\dot{{\cal H}} = 0$.

Expectation values can be used to simulate experiments, measuring induced properties (changes in expectation value)
as a function of time. Subsequent Fourier transformation of the signal provides direct spectral information without
resorting to time-dependent perturbation theory. A much-used example is the change in electric dipole moment induced
by an external electromagnetic field from which frequency-dependent polarizability and absorption spectrum
(after multiplication with suitable constants) can be obtained by Fourier transformation.

As an alternative to induced properties, quantum mechanical autocorrelation functions provide information about
energy levels and excitation energies directly from the state vectors at different points in time. A quantum mechanical autocorrelation
function is defined as the overlap (probability amplitude)
between state vectors at different times, see, e.g., Robinett's review on quantum wave packet revival~\cite{Robinett2004}
where autocorrelation functions play a central role.
In CC theory we can generalize the concept of autocorrelation function with the aid of the
indefinite inner product, Eq.~\eqref{eq:iip}, as
\begin{equation}
   A(t^\prime,t) \equiv \llangle S(t^\prime) \vert S(t)\rrangle
   .
\label{eq:autocorr}
\end{equation}
While the indefiniteness of the inner product implies that the absolute square of the CC autocorrelation function 
is bounded neither from below by $0$ nor from above by $1$, the
correct behavior is recovered in the FCI limit where the two terms in
Eq.~\eqref{eq:autocorr} are identical and give
$A(t^\prime,t)=\braket{\Psi(t^\prime)|\Psi(t)}$.

Note that if the Hamiltonian operator $H$ is independent of time and if the initial conditions correspond to the
system being in the ground state at time $t=0$, then $\tau_0(t) = -iE_0t$ with the ground-state energy given by
the usual projection formula $E_0 = \langle \Phi_0 \vert H \vert \psi \rangle$, and the other amplitudes
are constant. The energy $E_0$ is real if $H$, the orbitals, and
the amplitudes are all real. In this case we obtain $A(t^\prime,t) = \exp (-iE_0(t-t^\prime))$,
in agreement with exact quantum mechanics regardless of the CC truncation level.
In some situations, however, such as the presence of a static magnetic field,
the energy may attain an imaginary part, $E_0 = \langle \Phi_0 \vert H \vert \psi \rangle = x + iy$ with $x,y$ real.
In such cases, the autocorrelation function becomes
$A(t^\prime,t) = \exp (-ix(t-t^\prime)) \cosh (y(t-t^\prime))$, which deviates significantly from exact quantum mechanics when
$\cosh (y(t-t^\prime))$ is significantly different from $1$.

Two autocorrelation functions are of particular interest for the study
of the effects of short laser pulses on molecular electronic
systems. Assuming the electronic system is in the ground state at time
$t=0$, one relevant autocorrelation function is $A(0,t)$, which is the
probability amplitude of the system remaining in the ground state at a
later time $t>0$.  Assuming further that a laser pulse is active
in the time interval $t \in [0,t_1]$, the second interesting autocorrelation function is
$A(t_1,t)$, which is the probability amplitude of the the system
remaining (at time $t>t_1$) in the state created by the interaction
with the laser pulse.

To appreciate the information contained in $A(t_1,t)$, we will briefly
recapitulate its form in the FCI limit, which we may treat as exact. Let
$\{\ket{n}\}$ and $\{E_n\}$ denote the eigenstates and eigenvalues of
the field-free Hamiltonian. At time $t=t_1$ when the laser pulse is
turned off, the state of the system has evolved into the superposition
$\ket{\Psi(t_1)} = \sum_n \ket{n}c_n(t_1)$, where the complex
coefficients $c_n(t_1) = \braket{n | \Psi(t_1)}$ satisfy the
normalization condition $\sum_n \vert c_n(t_1) \vert^2 = 1$.  At later
times $t>t_1$, the state of the system is
$\ket{\Psi(t)} = \sum_n \ket{n}c_n(t_1)\exp (-iE_n(t-t_1))$ and the
autocorrelation function becomes
$A(t_1,t) = \sum_n \vert c_n(t_1)\vert^2 \exp (-iE_n(t-t_1))$.
Hence, Fourier transformation of $A(t_1,t)$ gives direct information
about the energy levels that contribute to the superposition at time
$t=t_1$ and their weight.
The autocorrelation function thus contains
essentially the same information as induced properties.

By analogy with exact quantum mechanics, we propose to use Eq.~\eqref{eq:autocorr} and its Fourier transform to analyze
explicitly time-dependent CC simulations. To further corroborate the proposal, we consider the behavior of
Eq.~\eqref{eq:autocorr} in the limit of weak external fields where the validity of perturbation theory can be assumed, at least
for the lowest-order corrections. Taking the ground state as the zeroth-order state and assuming that the zeroth-order
parameters are real, the autocorrelation function correct through first order in the perturbation is given by
\begin{align}
   A(t^\prime,t) &= e^{-\text{i}E_0(t-t^\prime)} \nonumber \\
        &\times 
        \left(1 + \text{i}\sum_{\mu\geq 0}\lambda_\mu^{(0)}\Im(\tau_\mu^{(1)}(t) - \tau_\mu^{(1)}(t^\prime))\right),
\label{eq:autocorr_pert}
\end{align}
where the superscripts $(0)$ and $(1)$ denote order.
It is well known from CC response theory~\cite{Koch1990} that the poles of the first-order amplitudes 
in the frequency domain can be interpreted as excitation energies of transitions between the ground state
and excited states. Fourier transformation of Eq.~\eqref{eq:autocorr_pert}
thus provides total energies due to the constant shift induced by the exponential prefactor $\exp (-\text{i}E_0(t-t^\prime))$.
An alternative justification can be constructed by linearization of $\bra{\tilde{\Psi}}$ and $\ket{\Psi}$ using the left
and right eigenvectors from equation-of-motion coupled-cluster (EOM-CC) theory.~\cite{Stanton1993,Krylov2008}

An essential difference between explicitly time-dependent theory and response theory is the absence of perturbation
expansions in the former, allowing for studies of electron dynamics in extreme environments where the fundamental assumptions
of perturbation theory are violated. The autocorrelation functions can be used to judge whether or not the system is in the
perturbative regime. If $\vert A(0,t) \vert^2$ is close to $1$, i.e., the electronic system largely remains in the ground
state, and if $\exp (-\text{i}E_0(t-t^\prime)) A(t^\prime,t)$ is close to $1$ with a small imaginary part,
then perturbation theory is likely valid.

\subsection{Integration of the time-dependent coupled-cluster equations}

The Hamiltonian structure of the time-dependent CC equations, and also
of the exact Schr\"odinger equation, implies that the theory of
symplectic, or canonical, transformations can be carried over from
classical mechanics.~\cite{Goldstein_CM} In particular, it follows
immediately that both the exact and approximate quantum mechanical
time evolutions are symplectic transformations. Clearly, preservation
of this property by a numerical integrator would be
beneficial.~\cite{HairerLubichWanner_GNI,Gray1996,Blanes2017}
Indeed, for such symplectic
integrators a backward error analysis can be done: there generally
exists, locally in time, a perturbed Hamiltonian system with
Hamilton function $\mathcal{H}_h = \mathcal{H} + O(h^p)$, where $p$ is the
order of the integrator and $h$ is the time step, which the integrator
solves \emph{excactly}. This implies that many physical conservation
laws are reproduced with a high degree of accuracy, such as
conservation of energy. Symplectic integrators typically also exhibit
long-time stability. Moreover, since a composition of symplecic maps
is again symplectic, integration of untruncated CC and the exact FCI
wavefunction would give equivalent results (assuming exact arithmetic)
using a symplectic integrator, as long as the exact state has non-zero
overlap with the chosen reference determinant.

A complication arises from the nonseparability of the Hamilton
function  ${\cal H}$ into a term depending only on $\tau$ and
another term depending only on $\lambda$. Symplectic integrators for
such nonseparable problems are generally implicit,
which means that the amplitudes at one time step can not be computed
from the amplitudes of the previous time step alone via an explicit
formula. 
Instead, a set of nonlinear equations must be solved iteratively with a computational complexity comparable to solving the ground-state
CC equations in every time step. Although, at first sight, this appears to make symplectic integration infeasible, the key parameter
in time-dependent CC simulations is the average number of evaluations of the Hamilton derivatives of Eq.~\eqref{eq:Hamilton}
per time step.
For the explicit RK4 integrator, which is not symplectic~\cite{HairerLubichWanner_GNI}
but often used for ordinary differential equations (ODEs),
four evaluations of both the Hamilton derivatives are required in each time step. In practice, even higher-order implicit symplectic
integrators may require significantly \emph{less} than four evaluations per time step on average, see, e.g., Table 6.1 of
Ref.~\onlinecite{HairerLubichWanner_GNI} for the Gauss integrator.

For notational convenience we collect the amplitudes $(\tau, \lambda)$ in the vector $y$ and write the time-dependent CC equations
as the ODE
\begin{equation}
    \dot{y} = f(y,t), \qquad y \in \mathbb{C}^{2m}, \quad f: \mathbb{C}^{2m}\times\mathbb{R} \mapsto \mathbb{C}^{2m},
\end{equation}
where $m$ is the number of amplitudes. With the initial condition $y(0) = y_0$ and discretizing time with a constant step $h$
such that $t_n = nh\ (n=0,1,2,\ldots)$, the RK4 integrator is defined by
\begin{align}
    &k_1 = f(y_n,t_n), \\
    &k_2 = f(y_n + \frac{h}{2}k_1, t_n + \frac{h}{2}), \\
    &k_3 = f(y_n + \frac{h}{2}k_2, t_n + \frac{h}{2}), \\
    &k_4 = f(y_n + hk_3,t_n + h),\\
    &y_{n+1} = y_n + \frac{h}{6} \left( k_1 + 2k_2 + 2k_3 + k_4 \right), 
\end{align}
where $y_n$ is the approximation of $y(t_n)$. This is a fourth-order explicit scheme, requiring exactly four $f$ evaluations per time step,
which is easily implemented given an implementation of the right-hand sides of Eqs.~\eqref{eq:rhs1} and~\eqref{eq:rhs2}.

A general implicit $s$-stage Runge--Kutta method is defined by 
\begin{align}
\label{eq:Gauss}
    &y_{n+1} = y_n + h\sum_{i=1}^s b_i f(y_n + Z_{in}, t_n + c_ih), \\
\label{eq:Gauss-Z}
    &Z_{in} = h\sum_{j=1}^s a_{ij} f(y_n + Z_{jn}, t_n + c_jh),
\end{align}
with real coefficients $a_{ij}, b_i, c_i,\
i,j=1,2,\ldots,s$. The Gauss integrator is a collocation method:
Interpolating the numerical solution between $t_n$ and
$t_n+h$ by a polynomial of order $s$ and requiring the ODE to be
satisfied at the $s$ Gauss--Legendre quadrature
points gives a symplectic and reversible integrator of order $2s$. The 
coefficients $b_i, c_i\ (i=1,2,\ldots,s)$ are abscissa and
weights, respectively, of the Gauss--Legendre quadrature, computed in our
implementation using the Golub--Welsch algorithm,~\cite{Golub1969}
and the matrix $a$ is then computed analytically (using the polynomial
antiderivative) from 
\begin{equation}
    a_{ij} = \int_0^{c_j} \ell_j(x) \text{d}x,
\end{equation}
where
\begin{equation}
    \ell_j(x) = \prod_{k=1, k\neq j}^s \frac{x - c_k}{c_j - c_k},
\end{equation}
is the $j$th Lagrange interpolation polynomial.
The Gauss integrator is implicit since the nonlinear equations~\eqref{eq:Gauss-Z} must be solved iteratively in each time step,
and the number of evaluations of $f$ thus depends on the number of iterations. Following the recommendations of
Hairer et al.~\cite{HairerLubichWanner_GNI} we use fixed-point iterations defined for $i=1,2,\ldots,s$ by
\begin{equation}
    Z_{in}^{(k+1)} = h\sum_{j=1}^s a_{ij}f(y_n + Z_{jn}^{(k)}, t_n + c_jh),
\end{equation}
where $k$ is the iteration counter. In our current implementation we have not used convergence acceleration techniques,
such as the Anderson acceleration for fixed-point iterations,~\cite{Walker2011} although this would most likely reduce
the number of evaluations, at least for larger time steps.

The initial guess is of crucial importance for rapid convergence of the fixed-point iterations and we have
implemented five different initial guesses. The two first are very simple; one, labeled 0,
consists of the guess $Z_{in}^{(0)} = 0$ and the other, labeled 1, consists of the guess
$Z_{in}^{(0)} = hc_if(y_n,t_n + c_ih)$. The remaining three initial guesses, labelled A, B, and C, require somewhat more
computation and are described (with the same
labels) in section VIII.6.1 of Ref.~\onlinecite{HairerLubichWanner_GNI}. Note, however, that our initial guess B is the
one of Ref.~\onlinecite{HairerLubichWanner_GNI} of order $s+1$ requiring a single additional $f$-evaluation. For initial guesses
1, A, and C we thus need at least $s$ evaluations per time step, while for initial guess B we need at least $s+1$
evaluations per time step. The best-case scenario is achieved when the initial guess solves the equations (to within a
given numerical threshold), making the fixed-point iterations converge after a single iteration.

\section{Numerical experiments}
\label{sec:numexp}

\subsection{Implementation notes and computational details}

An implementation of the time-dependent CC equations, in principle,
requires nothing more than a standard ground-state calculation of both
$\tau$ and $\lambda$ amplitudes. The major obstacle is that the code
must support complex algebra, thus conflicting with all highly
optimized open-source quantum-chemistry implementations of the CC
hierarchy of methods. Consequently, the core of our pilot
implementation of the time-dependent CC equations consists of a Python
code for automatic derivation of CC formulas using the SymPy
module,~\cite{sympy} combined with automatic code generation. All
required integrals over contracted Gaussian basis functions and
optimized HF orbitals are obtained through the NumPy interface of the
Psi4 open-source quantum chemistry program.~\cite{Parrish2017} While
our pilot implementation automatically factorizes tensor contractions
using intermediates to obtain optimal asymptotic scaling with respect
to the number of HF orbitals, the use of intermediates is not
optimized. Moreover, our implementation does not utilize point group
symmetries of small molecules. Thus, application is currently limited to
very small test cases.  This is sufficient for the scope of the
present work, however.  Note that our implementation is spin-unrestricted.

We write the electronic Hamiltonian in the semi-classical approximation as
\begin{equation}
   H = H(t) = H_0 + V(t),
\end{equation}
where
\begin{equation}
    H_0 = F + W,
\end{equation}
is the time-independent external field-free electronic Hamiltonian consisting of the sum of the Fock operator $F$ and the 
fluctuation potential $W$. Although not imperative for our impementation, we use canonical HF orbitals from Psi4
such that the Fock operator is diagonal. 
The interaction with an external uniform electric field is described by the operator
\begin{equation}
\label{eq:V}
   V(t) = -\boldsymbol{d}\cdot\boldsymbol{E} \cos (\omega (t-t_0)) G(t),
\end{equation}
where $\boldsymbol{d}$ is the electric-dipole operator, $\boldsymbol{E}$ is the constant electric field vector,
$\omega$ is the carrier frequency, $t_0$ the start time, and $G(t)$ is an envelope function controlling duration and
temporal shape of the interaction. We use either the sinusoidal envelope
\begin{equation}
\label{eq:sinusoidal_envelope}
   G(t) = \sin^2 \left( \frac{\pi (t-t_0)}{t_d} \right) \theta(t-t_0) \theta(t_d-(t-t_0)),
\end{equation}
where $t_d$ is the duration and $\theta(t)$ is the Heaviside step function,
or the Gaussian envelope
\begin{equation}
\label{eq:gaussian_envelope}
   G(t) = \exp \left( \frac{(t-t_c)^2}{2w^2}  \right),
\end{equation}
where $t_c$ is the center and $w$ the width of the Gaussian.

For simplicity, and due to the limitations of our pilot implementation, we use the He and Be atoms as test systems for 
time-dependent CCSD simulations. For He,
CCSD is equivalent to FCI, i.e. formally exact for the chosen basis set, whereas for Be the CCSD method is approximate.
In all cases, the CCSD ground state is used as the initial state $\vert S(0)\rrangle$.
All quantities except dipole moments are computed with normal-ordered operators and thus
contain the correlation contribution only.

For comparison we also run explicitly time-dependent FCI calculations on the He and Be atoms, using the FCI module 
of the PySCF software framework~\cite{PySCF}
and the Gauss integrator for propagation with the FCI ground state as initial state.
Further, we compute excited states using the (unrestricted) EOM-CCSD implementation of PySCF.

\subsection{Conservation of energy}

Capturing the correct physical behavior is the primary objective of an approximate integrator. The most direct
way to measure this is through conservation laws and, as a general example, we consider here the conservation of energy.
If we subject an electronic system to an external force in a finite time interval, the energy should be
constant before and after this interval (but not during the application of the external force).

\begin{figure}[!ht]
\begin{center}
\includegraphics[width=\columnwidth]{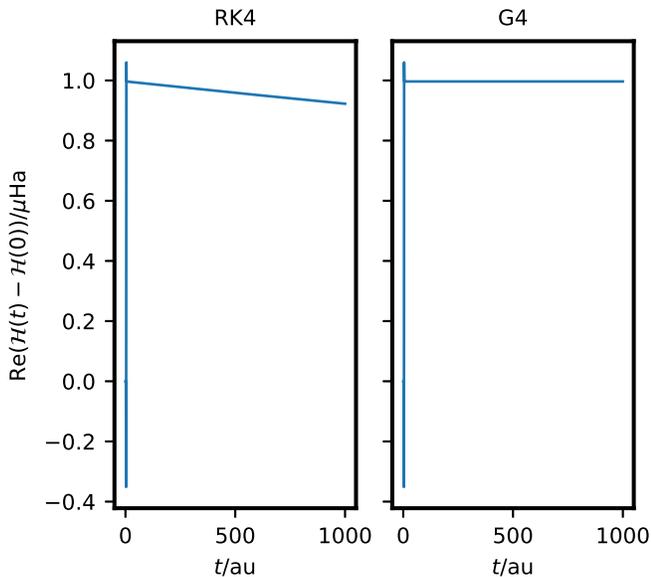}
\caption{\label{fig:energy_conservation} Total CCSD correlation energy computed as the real part of the Hamilton
function ${\cal H}$ as a function of time for the RK4 and fourth-order Gauss integrators for He with the cc-pVDZ basis set
and time step $h=0.1\, \text{au}$.
}
\end{center}
\end{figure}

As a test system we choose the He atom, for which the CCSD method is equivalent to FCI, using the cc-pVDZ basis
set.~\cite{Dunning1989}
We subject the He atom to an electric-field kick of strength $0.002\, \text{au}$ along the $z$-axis
defined by Eqs.~\eqref{eq:V} and \eqref{eq:gaussian_envelope} with the parameters $\omega = 0\, \text{au}$,
$t_c = 3\, \text{au}$, and $w = 0.5\,\text{au}$. The external force thus is applied for about $7$ atomic units of time.
The time-dependent CCSD equations are integrated using the RK4 and the fourth-order ($s=2$) Gauss integrator (denoted G4)
with start guess C for the fixed-point iterations in each time step.
The total simulation time is $1000\,\text{au}$ and the time step $h=0.1\, \text{au}$.
The correation energy computed at each time step as the real part of the Hamilton function ${\cal H}$ is plotted in
Fig.~\ref{fig:energy_conservation}.
While the Gauss integrator maintains a constant energy after the external force has been applied, the RK4
integrator leads to loss of energy, explicitly
demonstrating the benefit of applying a symplectic method.
Since CCSD is equivalent to FCI for this system,
the Hamilton function should be manifestly real, at least within exact
arithmetic.
\begin{figure}[!ht]
\begin{center}
\includegraphics[width=\columnwidth]{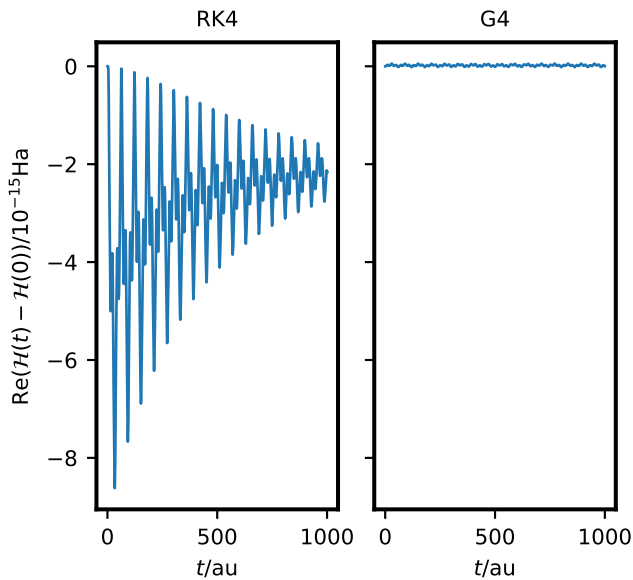}
\caption{\label{fig:energy_conservation_imag} The imaginary part of the CCSD Hamilton
function ${\cal H}$ as a function of time for the RK4 and fourth-order Gauss integrators for He with the cc-pVDZ basis set
and time step $h=0.1\, \text{au}$.
}
\end{center}
\end{figure}
The imaginary part is indeed very small with both integrators but, as
shown in Fig.~\ref{fig:energy_conservation_imag}, the RK4 integrator
leads to an oscillatory imaginary part, which is about two orders of
magnitude greater than that observed with the G4 integrator. The
maximum absolute imaginary part of the CCSD Hamilton function is
$9\cdot 10^{-15}\,\text{Ha}$ with the RK4 integrator compared with
$5\cdot 10^{-17}\,\text{Ha}$ with the G4 integrator.
In order to test whether these values are meaningful and not merely numerical
noise, we investigate the effect of increasing and decreasing the time step and find
that the ratio of about $100$ persists:
Doubling the time step to $h=0.2\,\text{au}$, the maximum absolute
imaginary part of the CCSD Hamilton function increases to $7\cdot 10^{-14}\,\text{Ha}$ with the
RK4 integrator and $5\cdot 10^{-16}\,\text{Ha}$  with the G4 integrator.
Halving the time step to $h=0.05\,\text{au}$, the maximum absolute 
imaginary part of the CCSD Hamilton function decreases to $3\cdot 10^{-16}\,\text{Ha}$ with the
RK4 integrator and $3\cdot 10^{-18}\,\text{Ha}$  with the G4 integrator.

While tiny, we
speculate that the imaginary parts may become
significantly greater for the larger time steps required for realistic
simulations on larger molecules, and that a symplectic method will
outperform RK4 in this respect. For example, running the same simulation for the slightly
larger Be atom with the cc-pVDZ basis, the maximum absolute imaginary part of
the CCSD Hamilton function increases to
$2\cdot 10^{-14}\,\text{Ha}$ with the RK4 integrator and
$2\cdot 10^{-16}\,\text{Ha}$ with the G4 integrator; see also
Ref.~\onlinecite{Pigg2012}.

We stress that although symplecticity is highly desirable, the RK4 integrator may still 
produce sufficiently accurate results for properties other than the energy.
For example, computing the lowest-lying singlet excitation energy ($1\text{s} \rightarrow 2\text{p}$ transition) through
the Fast Fourier Transform (FFT) of the dipole moment induced by the electric-field kick yields the same
result, $2.871\,\text{Ha}$, for He with the RK4 and G4 integrators.
This agrees with the excitation energy obtained from EOM-CCSD (diagonalization), $2.874\,\text{Ha}$,
to within the frequency resolution of the FFT, which is $0.006\,\text{Ha}$ in this case.
Moreover, the energy drift of the RK4 integrator may be minimized
by reducing the time step, albeit at the expense of computational cost due to the increasing number of $f$ evaluations
required.

\subsection{Computational performance of Gauss integrators}

We now turn to the efficiency of the Gauss integrator as measured by the number of $f$ evaluations per time step,
which should be compared with the four evaluations per time step required by the RK4 integrator. As discussed above,
the crucial parameter here is the initial guess for the fixed-point iterations. We use the same test system as in the
previous section, i.e., the He atom with the cc-pVDZ basis set exposed to the same electric-field kick. The number of
$f$ evaluations per time step is measured for simulations of duration $20\,\text{au}$.

\begin{figure}[!ht]
\begin{center}
\includegraphics[width=\columnwidth]{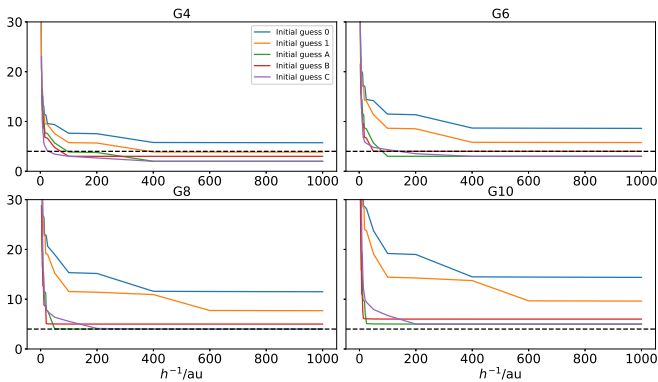}
\caption{\label{fig:f_eval_per_step} Number of $f$ evaluations per time step as a function of the inverse time step size
for Gauss integrators of orders $4, 6, 8, 10$ for He with the cc-pVDZ basis set and simulation time $20\,\text{au}$.
The horizontal dashed lines mark the number (four) of $f$ evaluations
per time step required by the RK4 integrator.
}
\end{center}
\end{figure}

Figure~\ref{fig:f_eval_per_step} shows the number of $f$ evaluations per time step required by the Gauss integrators of order
$4$, $6$, $8$, and $10$ as functions of $h^{-1}$, the number of steps taken to propagate through one atomic unit of time.
As discussed above, the number of $f$ evaluations per time step depends on the number of iterations required to converge
the fixed-point iterations and thus depends crucially on the initial guess for these iterations. 
It is immediately clear from Fig.~\ref{fig:f_eval_per_step} that all five initial guesses perform poorly with large
time steps. Although the performance is likely to improve by application of convergence acceleration techniques,
it will be difficult to compete with the RK4 integrator with larger time steps. On the other hand, this regime is where
the symplecticity problems of the RK4 integrator are most pronounced and the additional cost of the Gauss integrators may be
time well spent. As the time step is decreased, the number of $f$ evaluations decrease dramatically with the more advanced
initial guesses A, B, and C. With sufficiently small time steps, the implicit Gauss integrators 
require exactly $s$ (for initial guesses A and C) or $s+1$ (for initial guess B) $f$ evaluations per time step,
meaning that they are effectively equivalent
to explicit integrators in terms of computational effort. Remarkably, for smaller time steps and using initial guesses A or C, the Gauss
integrators up to and including order 8 are at least as fast as the RK4 integrator with the same time step.

\begin{figure}[!ht]
\begin{center}
\includegraphics[width=\columnwidth]{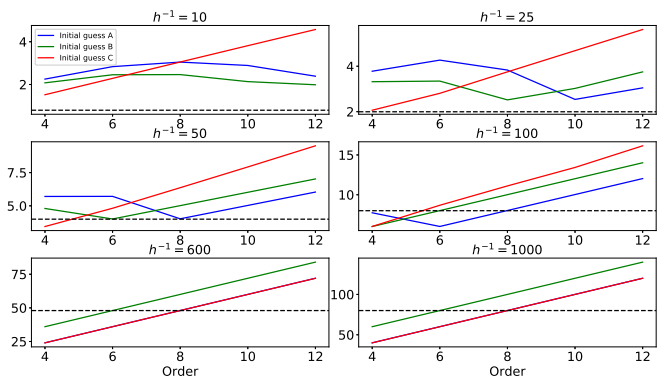}
\caption{\label{fig:f_eval_tot} Total number of $f$ evaluations (in thousands) as a function of the order of the Gauss integrator, plotted
as lines for visual aid, for He with the cc-pVDZ basis set and simulation time $20\,\text{au}$.
The horizontal dashed lines mark the number of $f$ evaluations used by the RK4 integrator with the indicated time step.
In the last two plots, the lines for initial guesses A and C coincide.
}
\end{center}
\end{figure}

These features are also evident from Fig.~\ref{fig:f_eval_tot}, which shows the total number of $f$ evaluations as a function of
the order of the Gauss integrator for the three initial guesses A, B, C.
Note, in particular, that initial guesses A and B may be more efficient at higher orders for larger steps,
since $s$ (half the order of the integrator), also determines the order of the initial guess.~\cite{HairerLubichWanner_GNI}
With $h=0.1\,\text{au}$, for example, initial guess B requires slightly fewer $f$ evaluations at order $12$
than at order $4$, and with $h=0.02\,\text{au}$, initial guess A reduces the effort by about $1/3$ at order $8$ compared
with order $4$, making the G8 integrator virtually indistinguishable from the RK4 integrator in terms of computer time.

At any rate, the total number of $f$ evaluations renders time-dependent CC simulations a formidable computational task
compared with ground-state and response calculations, which require on the order of $10$--$100$ evaluations.
Consequently, the application scope of time-dependent CC theory must be the study of processes where the quantum
dynamics is essential and/or where time-dependent perturbation theory breaks down.

\subsection{Autocorrelation functions}

We now investigate the autocorrelation function, Eq.~\eqref{eq:autocorr}, as a tool for extracting information about
stationary states in time-dependent CC simulations. To this end, we first expose the system to a laser pulse of
the form \eqref{eq:V} with the sinusoidal envelope \eqref{eq:sinusoidal_envelope} from time $t=0$ to time $t=t_1$.
Subsequently, we record the induced dipole moment and the autocorrelation function $A(t_1,t)$ for $t>t_1$.

We first expose the He atom to a laser pulse defined by Eqs.~\eqref{eq:V} and \eqref{eq:sinusoidal_envelope} with the
parameters $\omega = 2.8735643\,\text{au}$, $t_0 = 0\,\text{au}$, and $t_d = 5\,\text{au}$. The carrier frequency corresponds
to the lowest-lying EOM-CCSD/cc-pVDZ electric-dipole allowed transition from the ground state of He and the electric field is
linearly polarized along the $z$-axis. We then simulate the evolution of the electronic
system in two steps, first in the presence of the laser pulse from $t = t_0 = 0\,\text{au}$ to $t=t_1=t_d=5\,\text{au}$ and
then in the absence of a field from $t=t_1$ to $t=t_1+5000$ at the CCSD/cc-pVDZ level using the G6 integrator with time
step $h=0.01\,\text{au}$. The resulting frequency resolution is $1.26\cdot 10^{-3}\,\text{au}$.
This procedure is repeated with varying electric-field strengths ranging from $10^{-3}\,\text{au}$ to $10\,\text{au}$,
with corresponding ponderomotive energies\cite{Krausz2009} $U_{\text{p}} = \boldsymbol{E}^2
/ 4\omega^2$ in the range from $10^{-8}\,\text{au}$ to $3\,\text{au}$.
The field strengths thus range from very weak to slightly above the perturbation limit, which is taken to be the line in
an intensity--frequency plot where the ponderomotive energy is equal to the carrier frequency. We note in passing that none
of the field strengths are near the relativistic limit.

\begin{figure}[!ht]
\begin{center}
\includegraphics[width=\columnwidth]{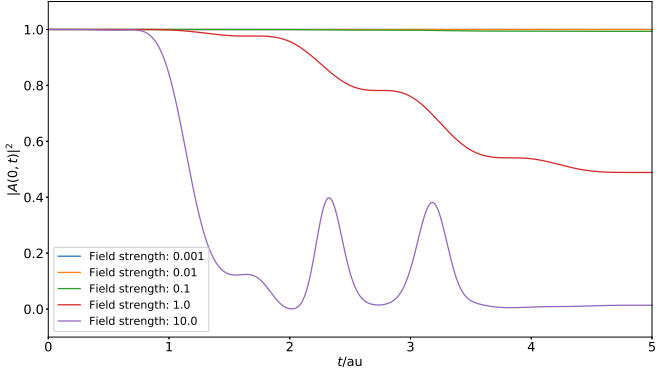}
\caption{\label{fig:he_gs_prob} Evolution of the ground-state probability during the laser pulse for the He atom at different
field strengths computed at the time-dependent CCSD/cc-pVDZ level of theory.
}
\end{center}
\end{figure}

\begin{figure}[!ht]
\begin{center}
\includegraphics[width=\columnwidth]{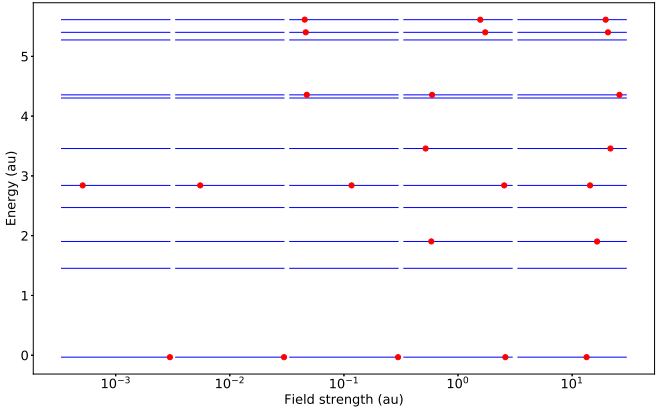}
\caption{\label{fig:he_levels} Unrestricted EOM-CCSD energy levels of He with the cc-pVDZ basis.
Circles indicate the
energy levels contributing to the CC state as detected by FFT of the autocorrelation function $A(t_1,t)$ at the given field strength.
The horizontal position of each circle indicates the relative and normalized weight of that level obtained from the peak intensities
of the FFT. The weight scale is logarithmic
and extends from $10^{-5}$ (left edge of each horizontal line) to $1$ (right edge of each horizontal line).
}
\end{center}
\end{figure}

\begin{figure}[!ht]
\begin{center}
\includegraphics[width=\columnwidth]{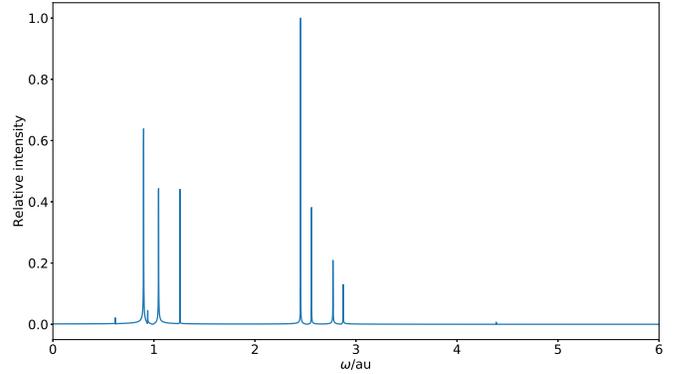}
\caption{\label{fig:he_dip_10} Dipole spectrum of He at field strength $10\,\text{au}$ 
computed at the time-dependent CCSD/cc-pVDZ level of
theory. 
}
\end{center}
\end{figure}

The ground-state probability for He during the laser pulse is plotted as a function of time in Figure~\ref{fig:he_gs_prob}
for each field strength.
These curves are identical to the probabilities computed with regular time-dependent FCI using the same integrator,
validating the time-dependent CCSD implementation.
The final ground-state probabilities (at $t=5\,\text{au}$), in order of increasing field strength,
are  $99.9999\,\%$, $99.9932\,\%$, $99.3213\,\%$, $48.8647\,\%$, and $1.3835\,\%$, 
showing that one of the fundamental assumptions of time-dependent perturbation theory---the ground-state probability
must remain close to unity---is valid for field strengths up to at least $10^{-1}\,\text{au}$ in this case.
The final state with field strength $10\,\text{au}$, on the other hand, is nearly orthogonal to the ground state.
At this field strength we also note that induced emission causes partial ``revival'' of the ground state while the laser pulse is on.
As argued above, Fourier transformation of the autocorrelation function $A(t_1,t)$ ($t_1 = 5\,\text{au}$) provides
information of about the populated excited energy levels. This information is plotted in Fig.~\ref{fig:he_levels}
superimposed on the energy levels obtained from EOM-CCSD.
The weights plotted in this figure are indicative of, but not identical to, the probabilities associated with each energy level,
since the former are computed from renormalized relative peak intensities of the FFT of the signal. At the lowest field strengths
only one excited level is populated and, indeed, only the $0 \to 4$
transition is observed in the dipole spectrum computed
by FFT of the induced dipole moment at $t \in
[5,5005]\,\text{au}$. (The states are numbered according to increasing
energy.) Higher-lying levels become weakly populated at field strength
$0.1\,\text{au}$ and the dipole spectrum now contains very weak lines that can be assigned to transitions between excited states
(in order of decreasing intensity: 
$4$ and $10$, $4$ and $9$, $7$ and $10$, $7$ and $9$).
Going to field strength $1\,\text{au}$, the dipole-forbidden (i.e., no direct electric-dipole transition from the ground state)
level $2$ becomes populated and the weight of level $4$ rivals that of the ground state, which is still the most probable
stationary state contributing to the CC state and the transition $0 \to 4$ remains the most intense in the dipole spectrum.
This is no longer the case at field strength $10\,\text{au}$ where the ground state
has become the least likely, as one might also suspect from its low probability $\vert A(0,t_1) \vert^2 = 1.3835\,\%$.
The CCSD dipole spectrum at field strength $10\,\text{au}$ is plotted in Fig.~\ref{fig:he_dip_10}.
The most intense line corresponds to the transition $2 \to 7$ and the transition $0 \to 4$, which is the most intense line
at the lower field strengths, has dropped to a relative intensity of $0.13$.

\begin{figure}[!ht]
\begin{center}
\includegraphics[width=\columnwidth]{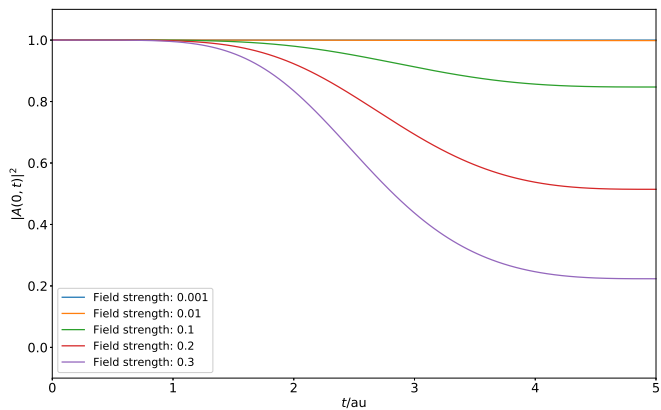}
\caption{\label{fig:be_gs_prob} Evolution of the ground-state probability during the laser pulse for the Be atom at different
field strengths computed at the time-dependent CCSD/cc-pVDZ level of theory.
}
\end{center}
\end{figure}

Running the same simulations for the Be atom with the cc-pVDZ basis set at field strengths
ranging from $10^{-3}\,\text{au}$ to $0.3\,\text{au}$ and carrier frequency $\omega = 0.2068175\,\text{au}$,
which corresponds to the lowest-lying EOM-CCSD/cc-pVDZ electric-dipole allowed transition from the ground state,
leads to the ground-state probabilities during the laser pulse plotted in Fig.~\ref{fig:be_gs_prob}.
Despite the lower field strengths compared with the He case discussed above, the lower carrier frequency ensures comparable
ponderomotive energies ranging from $5.84 \cdot 10^{-6}\,\text{au}$ to $0.53\,\text{au}$, the latter being well above the
perturbative limit. The lower carrier frequency also means that the laser pulse is less oscillatory, leading to more
monotonic decay of the ground-state probability. The final probabilities are $99.998\,\%$, $99.835\,\%$, $84.728\,\%$,
$51.440\,\%$, and $22.331\,\%$, clearly indicating the deviation from a perturbative treatment at the greater field strengths.
Running the same simulation at the time-dependent FCI/cc-pVDZ level with field strength $0.3\,\text{au}$ yields a ground-state
probability curve virtually indistuinguishable from the CCSD one in Fig.~\ref{fig:be_gs_prob}: the root-mean-square deviation
between the curves is merely $3 \cdot 10^{-4}$ with a maximum deviation of $4 \cdot 10^{-4}$.

\begin{figure}[!ht]
\begin{center}
\includegraphics[width=\columnwidth]{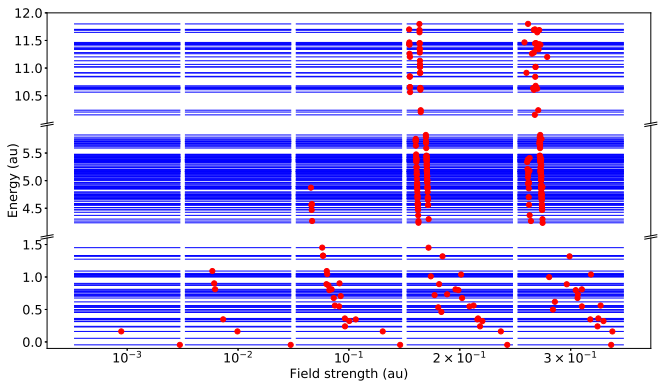}
\caption{\label{fig:be_levels} Unrestricted EOM-CCSD energy levels of Be with the cc-pVDZ basis.
Circles indicate the
energy levels contributing to the CC state as detected by FFT of the autocorrelation function $A(t_1,t)$ at the given field strength.
The horizontal position of each circle indicates the relative and normalized weight of that level obtained from the peak intensities
of the FFT. The weight scale is logarithmic
and extends from $10^{-5}$ (left edge of each horizontal line) to $1$ (right edge of each horizontal line).
}
\end{center}
\end{figure}

\begin{figure}[!ht]
\begin{center}
\includegraphics[width=\columnwidth]{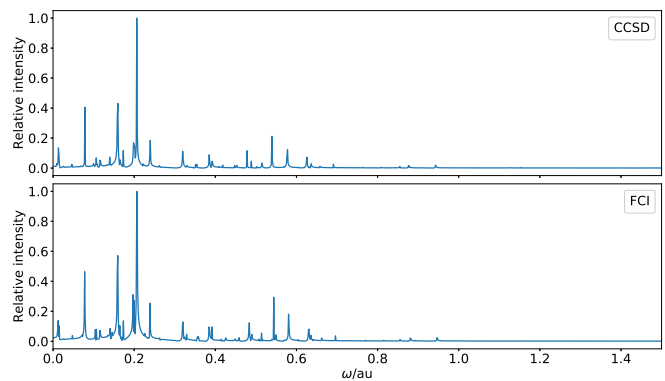}
\caption{\label{fig:be_dip_0_3} Dipole spectrum of Be at field strength $0.3\,\text{au}$ 
computed at the time-dependent CCSD/cc-pVDZ and FCI/cc-pVDZ levels of theory.
}
\end{center}
\end{figure}

The contributing energy levels computed by FFT of the autocorrelation function $A(t_1,t)$ are superimposed on the EOM-CCSD
levels in Fig.~\ref{fig:be_levels}.
At field strength $10^{-3}\,\text{au}$ the only contributing excited state is the lowest-lying dipole-allowed state that
can be reached from the ground state. Five excited states contribute at field strength $10^{-2}\,\text{au}$ and analysis of the dipole
spectrum reveals that all five states are reached by direct excitation from the ground state.
Transitions between excited states appear at field strength $0.1\,\text{au}$, although specific assignment of each line
in the dipole spectrum is made difficult by the limited frequency resolution (caused by finite simulation time)
combined with the closeness of the excited states. More than $70\,\%$ of the states contribute at field strengths
$0.2\,\text{au}$ and $0.3\,\text{au}$, albeit mostly with very low weight. At field strength $0.3\,\text{au}$ the ground state
is no longer the most probable state and several excited levels have comparable weights. The dipole spectrum,
shown in Fig.~\ref{fig:be_dip_0_3} along with the spectrum computed at the FCI level,
clearly displays several transitions between close-lying excited states.
The most intense
line peaks at $0.206\,\text{au}$ and corresponds to the $0 \to 2$ transition, which is the lowest-lying electric-dipole allowed
transition from the ground state. Aside from minor differences in relative intensities, the CCSD spectrum agrees very well with
the FCI spectrum, increasing the confidence in our time-dependent CCSD implementation.

Equation~\eqref{eq:autocorr_pert} offers an alternative assessment of the sufficiency of a perturbative treatment.
Table~\ref{tab:A_pert} reports the root-mean-square deviation from $1$ and $0$ of the real and imaginary parts
of $\tilde{A}(t_1,t) = \exp(\text{i}E_0(t-t_1))A(t_1,t)$, respectively, for He and Be.
\begin{table}[!ht]
\caption{\label{tab:A_pert}
Root-mean-square deviation of $\tilde{A}(t_1,t) = \exp(\text{i}E_0(t-t_1))A(t_1,t)$
computed at the time-dependent CCSD/cc-pVDZ level of theory for $t_1=5\,\text{au}, t \in [5,5005]\,\text{au}$.
}
\begin{scalebox}{0.95}{
\begin{tabular}{llllll}
\hline\hline
\multicolumn{6}{l}{He} \\
Field strength (au) & $10^{-3}$ & $10^{-2}$ & $10^{-1}$ & $10^{0}$ & $10^{1}$ \\
Re$[\tilde{A}(t_1,t)]$ & $8 \cdot 10^{-7}$ & $8 \cdot 10^{-5}$ & $8 \cdot 10^{-3}$ & $6 \cdot 10^{-1}$ & $1$ \\
Im$[\tilde{A}(t_1,t)]$ & $1 \cdot 10^{-4}$ & $1 \cdot 10^{-4}$ & $5 \cdot 10^{-3}$ & $3 \cdot 10^{-1}$ & $4 \cdot 10^{-1}$ \\
\hline
\multicolumn{6}{l}{Be} \\
Field strength (au) & $10^{-3}$ & $10^{-2}$ & $10^{-1}$ & $2 \cdot 10^{-1}$ & $3 \cdot 10^{-1}$ \\
Re$[\tilde{A}(t_1,t)]$ & $2 \cdot 10^{-5}$ & $2 \cdot 10^{-3}$ & $2 \cdot 10^{-1}$ & $5 \cdot 10^{-1}$ & $8 \cdot 10^{-1}$ \\
Im$[\tilde{A}(t_1,t)]$ & $4 \cdot 10^{-4}$ & $1 \cdot 10^{-3}$ & $1 \cdot 10^{-1}$ & $2 \cdot 10^{-1}$ & $2 \cdot 10^{-1}$ \\
\hline\hline
\end{tabular}
}\end{scalebox}
\end{table}
Overall, the root-mean-square deviations agree with the results above. For He, field strengths $1\,\text{au}$ and $10\,\text{au}$
are clearly non-perturbative, and for Be, field strengths $0.1$--$0.3\,\text{au}$ are non-perturbative.

\subsection{Very strong fields}

Increasing the field strength beyond those of the previous section poses a numerical challenge to the CCSD model, even for the He
atom where it is formally equivalent to FCI. This can be understood from the following simple qualitative analysis.
As the field strength is increased, the ground-state probability tends to zero and since the ground state of He (and Be)
is vastly dominated by the HF reference determinant, we may take this to mean that the state of the system is a superposition
of excited determinants.
In the limit of untruncated cluster operators, $\vert \Psi\rangle$ and
$\langle \tilde{\Psi}\vert$ are proportional to the FCI state and its conjugate, respectively. As the FCI coefficient $C_0$
of the HF determinant approaches zero, the amplitudes of the CCSD state must behave in a rather extreme fashion:
\begin{align}
   &C_0 = \langle \Psi \vert \Psi\rangle^{-1/2} e^{iy}e^{x} \to 0, \\
   &C_0^* = \langle \tilde{\Psi}\vert\tilde{\Psi}\rangle^{-1/2} e^{-iy}e^{-x} \nonumber \\
   &\qquad \times  \left( 1 - \lambda_1 \tau_1 - \lambda_2\tau_2 + \frac{1}{2}\lambda_2\tau_1^2) \right) \to 0,
\end{align}
where $\lambda_i, \tau_i, i=1,2$, collectively denote the suitable singles and doubles amplitudes, and
\begin{align}
   &x(t) = \Re \tau_0(t) = \Im \int_0^t \langle \Phi_0 \vert e^{-T(t^\prime)}H(t^\prime)e^{T(t^\prime)} \vert \Phi_0\rangle\,\text{d}t^\prime, \\
   &y(t) = \Im \tau_0(t) = -\Re \int_0^t \langle \Phi_0 \vert e^{-T(t^\prime)}H(t^\prime)e^{T(t^\prime)} \vert \Phi_0\rangle\,\text{d}t^\prime.
\end{align}
Evidently, the absolute value of the $\tau$ amplitudes must increase to make $\langle \Psi \vert \Psi\rangle$ and
$\langle \tilde{\Psi}\vert\tilde{\Psi}\rangle$ as large as possible while maintaining the absolute value of $x$ low enough to overcome
the exponential increase of either $\exp (x)$ or $\exp (-x)$. By the same token, the $\lambda$ amplitudes must not increase too
much. This represents a delicate numerical challenge.

\begin{figure}[!ht]
\begin{center}
\includegraphics[width=\columnwidth]{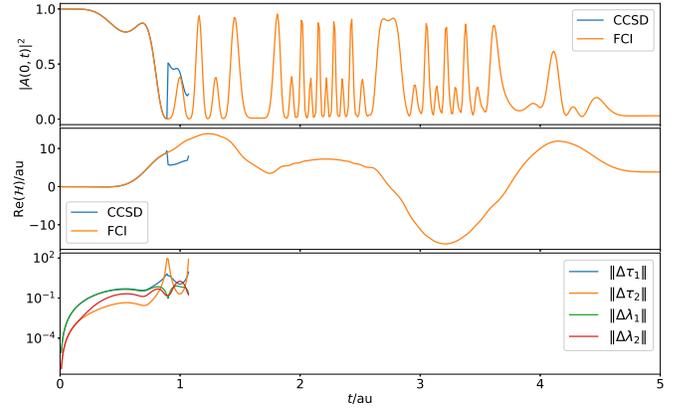}
\caption{\label{fig:he_fail} Time-dependent CCSD and FCI simulations of He with the cc-pVDZ basis exposed to a laser
pulse of field strength $100\,\text{au}$ and carrier frequency $\omega=2.8735643\,\text{au}$. Top panel: ground-state
probability. Middle panel: real part of the Hamilton function. Bottom panel: norm of the change in amplitudes relative to
the initial (ground) state.
}
\end{center}
\end{figure}

An example is presented in Fig.~\ref{fig:he_fail} for the He atom when the field strength is increased to
$100\,\text{au}$---all other parameters of the sinusoidal laser pulse and the G6 integrator
are the same as in the previous section. This corresponds to a ponderomotive
energy of $303\,\text{au}$, which is equivalent to about $105$ photons at the carrier frequency.
The CCSD ground-state probability and Hamilton function are indistinguishable
from the FCI simulation until $t=0.88\,\text{au}$; at $t=1.07\,\text{au}$ the CCSD simulation fails
due to numerical instability.
The ground-state probability in this time interval drops to about $0.2\,\%$ accompanied by
rapid increase of (the norm of) the amplitudes, especially the $\tau$ amplitudes, which causes numerical instability
in double-precision arithmetic while solving Eq.~\eqref{eq:Gauss-Z}.

\begin{figure}[!ht]
\begin{center}
\includegraphics[width=\columnwidth]{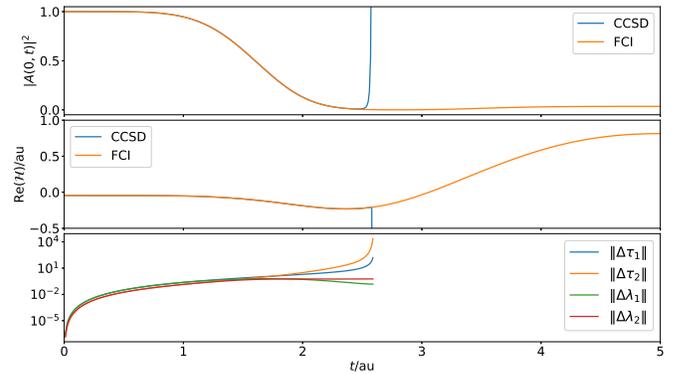}
\caption{\label{fig:be_fail} Time-dependent CCSD and FCI simulations of Be with the cc-pVDZ basis exposed to a laser
pulse of field strength $1\,\text{au}$ and carrier frequency $\omega=0.2068175\,\text{au}$. Top panel: ground-state
probability. Middle panel: real part of the Hamilton function. Bottom panel: norm of the change in amplitudes relative to
the initial (ground) state.
}
\end{center}
\end{figure}

As shown in Fig.~\ref{fig:be_fail},
the same problem appears in the simulation of the Be atom exposed to a laser pulse of $1\,\text{au}$, correponding to a ponderomotive
energy of $5.84\,\text{au}$ (roughly $28$ photons at the carrier frequency).
Again, the numerical instability can be traced to rapid changes in the amplitudes relative to the ground state as the ground-state
probability approaches zero.
In this case, CCSD is an approximation
and both the ground-state probility and Hamilton function differ from the FCI results, although the differences are too small to
be visible on the scale of the plots in Fig.~\ref{fig:be_fail}.

\begin{figure}[!ht]
\begin{center}
\includegraphics[width=\columnwidth]{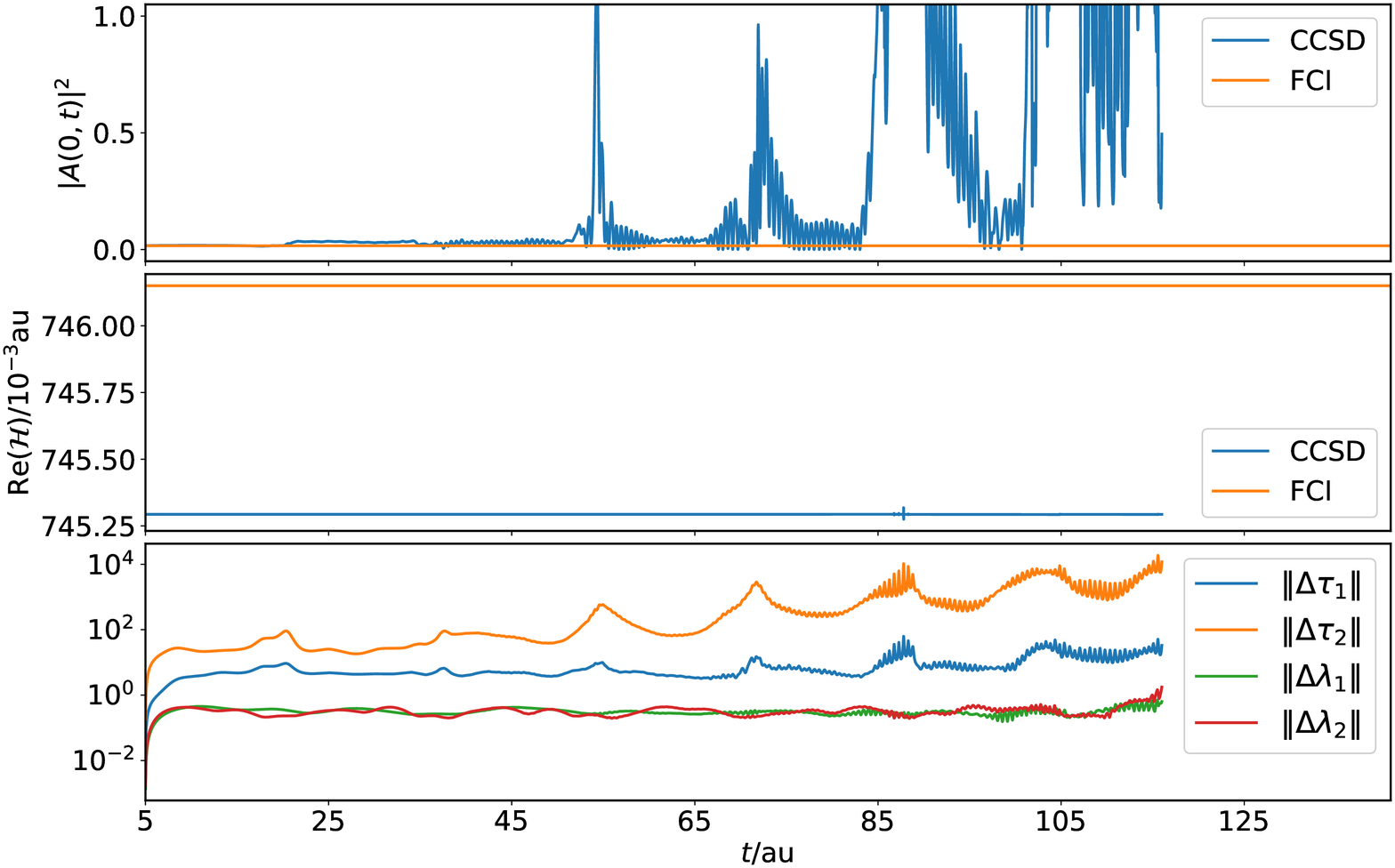}
\caption{\label{fig:be_fail2} Time-dependent CCSD and FCI simulations of Be with the cc-pVDZ basis exposed to a laser
pulse of field strength $0.5\,\text{au}$ and carrier frequency $\omega=0.2068175\,\text{au}$. Top panel: ground-state
probability. Middle panel: real part of the Hamilton function. Bottom panel: norm of the change in amplitudes relative to
the state at $t=5\,\text{au}$.
}
\end{center}
\end{figure}

Such numerical instabilities may also be encountered within the CCSD model at less intense field strengths, as illustrated 
by CCSD and FCI simulations of the Be atom with a field strength of $0.5\,\text{au}$ presented in Fig.~\ref{fig:be_fail2},
which shows results for $t>5\,\text{au}$---i.e., after the laser pulse has been turned off. At $t=5\,\text{au}$ the amplitude norms are roughly $3$ and $19$ for $\tau_1$ and $\tau_2$,
and $0.3$ and $0.6$ for $\lambda_1$ and $\lambda_2$. The norm of the change in amplitudes reported in Fig.~\ref{fig:be_fail2} are measured 
relative to these.
In agreement with the FCI value of $1.6\%$ at $t=5\,\text{au}$, the CCSD ground-state probability is $1.7\%$.
At times $t>5\,\text{au}$ both the ground-state probability and the Hamilton function should remain constant. This is indeed the case
for the FCI simulation but the CCSD method fails spectacularly for the ground-state probability at $t\gtrapprox 20\,\text{au}$.
This is caused by numerical noise accumulating in the amplitudes, eventually causing the simulation to fail.
Note, however, that the CCSD Hamilton function remains almost constant throughout, except for a sub-millihartree oscillation at
$t\approx 90\,\text{au}$. This feature can be ascribed to the symplecticity of the Gauss integrator.

One might speculate that this is an effect of a too small basis set, but repeating the simulations with the aug-cc-pVDZ and cc-pVTZ
basis sets leads to essentially identical behavior for Be with field strength $0.5\,\text{au}$.
No problems are observed with a minimal basis set, however, indicating that increasing the basis set further,
for example by inclusion of low-lying continuum functions to support ionization processes, is unlikely to resolve the
problem.

\section{Concluding remarks}
\label{sec:conclusion}

In this study we have
\begin{itemize}
\item exploited the Hamiltonian structure of time-dependent CC theory to propose the Gauss integrator as a stable algorithm
for solving the time-dependent CC equations,
\item proposed autocorrelation functions based on an indefinite inner product for analyzing the time-dependent CC state,
\item presented a pilot implementation and validated it through simulations of the He and Be atoms in short laser pulses
with increasing intensity, comparing with results obtained with the time-dependent FCI method, and
\item observed that the CCSD approach fails for very strong laser pulses due to numerically intractable increase in
the $\tau$ amplitudes as the ground-state probability approaches zero.
\end{itemize}
We stress that the CCSD failure persists in the FCI limit even if, mathematically, the combined exponential ($\tau$) and linear
($\lambda$) parametrization should be sufficiently flexible to fully describe the electron dynamics.
While it may be possible to devise an integrator with sufficient numerical stability,
possibly in conjunction with the use of even smaller time steps,
the most likely solution is to allow the underlying orbitals to participate in the correlated electron dynamics in a manner
similar to the approaches of Kvaal~\cite{Kvaal2012} or Sato et al.~\cite{Sato2018} but in such a way that the FCI limit is recovered for more than two
electrons.~\cite{Kohn2005,Myhre2018,Kvaal2012}
A properly constructed moving reference determinant would capture the main effects of the laser pulse (which is represented
by a one-electron operator), ensuring a significant overlap with the FCI wave function and thus leading to well-behaved
amplitudes.

\begin{acknowledgments}
This work was supported by the Research Council of Norway (RCN) through its Centres of Excellence scheme, project number 262695,
by the RCN Research Grant No. 240698,
and by the European Research Council under the European Union Seventh
Framework Program through the Starting Grant BIVAQUM, ERC-STG-2014 grant
agreement No 639508. 
Support from the Norwegian Supercomputing Program (NOTUR) through a grant of computer time (Grant No.\ NN4654K) is
gratefully acknowledged.
The authors thank Prof. C. Lubich for helpful discussions.
\end{acknowledgments}

\bibliography{paper}
\bibliographystyle{aipnum4-1}

\end{document}